\documentstyle[sprocl]{article} 

\def\Journal#1#2#3#4{{#1} {\bf #2}, #3 (#4)} 
 
\def\NPB{{\em Nucl. Phys.} B} 
\def\PLB{{\em Phys. Lett.}  B} 
 
\def\PRD{{\em Phys. Rev.} D} 
 
\def\be{\begin{equation}} 
\def\ee{\end{equation}} 
\def\bea{\begin{eqnarray}} 
\def\eea{\end{eqnarray}} 

\def\beq{\begin{equation}}
\def\eeq{\end{equation}}
\def\eq{\end{equation}}
\def\bea{\begin{array}}
\def\ea{\end{array}}

\def\drawbox#1#2{\hrule height#2pt 
        \hbox{\vrule width#2pt height#1pt \kern#1pt 
               \vrule width#2pt}
               \hrule height#2pt}

\def\Fund#1#2{\vcenter{\vbox{\drawbox{#1}{#2}}}}
\def\Asym#1#2{\vcenter{\vbox{\drawbox{#1}{#2}
              \kern-#2pt       
              \drawbox{#1}{#2}}}}
\def\Asymt#1#2{\vcenter{\vbox{\drawbox{#1}{#2}
              \kern-#2pt       
              \drawbox{#1}{#2}
              \kern-#2pt       
              \drawbox{#1}{#2}}}}
 
\def\fund{\Fund{6.5}{0.4}}
\def\asym{\Asym{6.5}{0.4}}

\def\sym{\Fund{6.5}{0.4} \kern-.5pt \Fund{6.5}{0.4}}
\def\symthree{\Fund{6.5}{0.4} \kern-.5pt \Fund{6.5}{0.4}
               \kern-.5pt \Fund{6.5}{0.4} }

\begin{document} 

\title{
\hfill {\normalsize SU-ITP 97-32} \\
\hfill {\normalsize hep-ph/9801007 } \\
\vspace{.25in}
Recent Developments in Dynamical Supersymmetry 
Breaking}

 

\author{Scott Thomas } 
 
\address{Physics Department, 
Stanford University,  
Stanford, CA 94305}



 
 
 
\maketitle\abstracts{
Some formal aspects of supersymmetry breaking are reviewed. 
The classic ``requirements'' for supersymmetry breaking include
chiral matter, a dynamical superpotential, and a classical 
superpotential which completely lifts the moduli space.
These ``requirements'' may be evaded in theories with
large matter representations.
The mechanisms of supersymmetry breaking by confinement and 
quantum deformation of the moduli space are explained,
with emphasis on the importance of identifying the relevant
degrees of freedom in the ground state. 
Supersymmetry breaking and the behavior of the Witten index
in non-chiral theories are discussed. 
The quantum removal of directions which are classically
unlifted is also illustrated.
Examples of product gauge group theories that admit dual 
descriptions of the non-supersymmetric ground state are also
presented. 
}

\section{Introduction}

If supersymmetry is a symmetry of nature, it is clearly broken
in the ground state. 
Understanding the mechanisms by which supersymmetry 
may be spontaneously broken is crucial in deciphering 
what types of supersymmetric theories may describe nature. 
Recently there has been considerable progress in 
more formal 
aspects of supersymmetry breaking. 
Most of these have followed from the recent revolution in our 
understanding of strong non-perturbative dynamics 
in $N=1$ supersymmetric gauge theories.\cite{review}
In this paper I review recent developments 
in understanding the formal requirements for, 
and mechanisms of, supersymmetry breaking.\cite{phenrev} 

In the next section the motivation for studying dynamical
supersymmetry breaking by non-perturbative gauge dynamics
is reviewed.  
The importance of identifying the relevant degrees of freedom
in the ground
state is emphasized.
The classic ``requirements'' for dynamical supersymmetry breaking
are reviewed, and the general 
means by which these may be evaded are discussed. 
An example of breaking by the mechanism of a dynamical superpotential
over a classical moduli space is also given.  
In \S 3 the mechanisms of supersymmetry breaking by confinement 
and quantum deformation of the moduli space are reviewed. 
Supersymmetry is broken in these cases even though the exact
non-perturbative superpotential vanishes in the absence
of a tree level superpotential. 
Non-chiral theories which break supersymmetry are discussed
in \S 4.
The manner in which the Witten index can vanish in such 
theories is explained. 
Finally, in \S 5 product gauge group theories are given 
which admit two dual descriptions of the non-supersymmetric ground
state.  
As a function of the parameters of the theory,
these dual descriptions do not have overlapping regions of 
applicability.


\section{Dynamical Supersymmetry Breaking}

The Witten index for a supersymmetric theory, 
${\rm Tr}(-1)^F$, counts the number of supersymmetric ground 
states.\cite{index}
This index is not modified at any order in perturbation theory. 
So supersymmetry is either broken classically at tree level,
or by non-perturbative dynamics.  
In nature, if supersymmetry has any relevance to the hierarchy 
problem, the supersymmetry breaking scale is certainly well below 
the Planck scale. 
This would not be the case with tree level breaking in the 
absence of very small parameters.
However, in theories which exhibit  
dimensional transmutation (such as asymptotically free
non-Abelian gauge theories) 
the dynamical scale can naturally
be hierarchically smaller than any fundamental scale. 

Since the scale for supersymmetry breaking is well below the 
Planck scale, 
one can hope that the breaking admits
an effective field theory description in the rigid supersymmetric
limit. 
This is not necessarily the case if some fields have Planck scale
expectation values, in which case supergravity effects can become
important.  
In addition, in the global limit we give up any hope of 
understanding the smallness of the cosmological constant. 
However, even in the local context, it is non-perturbative
field theory dynamics which is believed to break supersymmetry. 
For all these reasons it useful to study supersymmetry 
breaking in global non-Abelian gauge 
theories.




\subsection{Identifying the Relevant Degrees of Freedom}

\label{reldof}

Since the Hamiltonian is related to the supercharge, 
$H = {1 \over 2} \{ Q, Q^{\dagger} \}$
for translationally invariant states, 
supersymmetry breaking implies 
non-zero vacuum energy in the rigid limit,
$
Q^{\alpha} | 0 \rangle \neq 0 \Rightarrow  
H|0 \rangle \neq 0
$.
Stable supersymmetry breaking therefore requires a ground
state with non-zero energy.
The potential is an incoherent sum of $D$- and $F$-terms 
\beq
V = {1 \over 2} g^2 D^a D^a + 
   F_{\phi} K^{\phi \phi^*} F_{\phi}^*
\eq
where $D^a = K_{\phi} T^a \phi$ and 
$F_{\phi} = \partial_{\phi} W$. 
Note that the $D$- and $F$-terms can not interfere.
In order to check for a supersymmetric ground state with vanishing 
energy, it is therefore sufficient to consider the $F$-terms on the 
subspace of the full field space for which $D^a=0$.
The question of supersymmetry breaking may then be 
reduced to whether or not all the auxiliary equations of motion can 
be simultaneously satisfied with $F=0$ on the 
$D$-flat moduli space. 

The moduli space of $D$-flat directions 
is special for a number of reasons. 
First, the gauge symmetry generally 
exhibits a definite pattern of breaking, $G \rightarrow H$,
on the moduli space. 
Second, the microscopic fields of the theory break up into
heavy fields which are eaten by the super-Higgs mechanism, plus
light fields which parameterize $D^a=0$. 
These parameters are related to the ring of gauge invariant 
chiral operators, $\{ X_i \}$. 
It has recently been shown that 
the classical moduli space of $D$-flat directions is in 
one to one correspondence with the classical 
chiral ring.\cite{luty}
This considerably simplifies the analysis of gauge theories
since the set of gauge invariant operators plus relations is 
typically
much easier to find than an explicit parameterization of 
the $D$-flat directions. 

It is important to note that 
with broken supersymmetry the fields need not sit 
precisely on a $D$-flat direction in the ground state. 
However, physically for small Yukawa couplings, $\lambda \ll g$, 
the excitations taking the system away from $D^a=0$ are 
typically very heavy and can be integrated out, leaving the 
light moduli as effective degrees of freedom.  

In a theory with a supersymmetric ground state, much of
the power in analyzing strong dynamics comes from 
the holomorphy and $U(1)_R$ symmetry of the superpotential.  
These constraints, plus global symmetries and limits are often enough
to uniquely fix the full non-perturbative superpotential. 
However, in theories in which supersymmetry is broken, a 
quantitative description of the ground state and excited 
spectrum also requires knowledge of the K\"{a}hler potential. 
Even though the K\"{a}hler potential is not protected by any
non-renormalization theorems, its leading
behavior can be determined in 
two important limits.  
The first is for large expectation values along the
$D$-flat directions.  
In this limit the gauge group is highly Higgsed, and 
weak at the scale of the expectation values. 
The relevant K\"{a}hler potential is then just the 
classical canonical one for the microscopic fields,
$\phi$, projected
onto the $D$-flat directions
\beq
K = \left. \phi^{\dagger} \phi \right|_{D^a=0} ~~~~~~~ 
    \phi \gg \Lambda
\label{Kclass}
\eeq
The K\"{a}hler potential (\ref{Kclass}) receives small, calculable, 
quantum corrections in this limit. 
This is the standard limit in which all theories of global
supersymmetry breaking have been analyzed in the past. 

For small expectation values, $\phi \ll \Lambda$, the gauge
group is strongly coupled, and the K\"{a}hler potential in general
receives large uncalculable corrections. 
However, the dominant contribution to the K\"{a}hler is 
sometimes calculable at strong coupling.  
This occurs at points of enhanced global symmetry. 
Classically the gauge group is typically enhanced at such 
points, and the classical K\"{a}hler potential for the moduli 
becomes singular. 
Quantum mechanically however, 
non-perturbative degrees of freedom, $\varphi$,
(including confined and magnetic chiral multiplets and magnetic
gauge multiplets)
often become massless at enhanced symmetry points in order
to saturate global anomalies. 
In terms of these degrees of freedom the K\"{a}hler potential is 
smooth. 
If the quantum theory at the enhanced symmetry point is infrared free,
then in terms of the non-perturbative degrees of freedom
\beq
K ( \varphi=0)= \varphi^{\dagger} \varphi   
\eq

%

This non-perturbative information about the K\"{a}hler potential
is crucial in analyzing non-supersymmetric ground states
at strong coupling. 
For the mechanisms of supersymmetry breaking 
discussed in \S \ref{Wzero}
which do not rely on a dynamically generated superpotential, in a 
certain sense 
the information that supersymmetry is in fact  
broken is contained in the K\"{a}hler potential. 
At strong coupling it is therefore very important to 
identify the relevant non-perturbative degrees of freedom.

\subsection{Classic Requirements for Supersymmetry Breaking}

As discussed in \S \ref{reldof} supersymmetry is broken
if all the auxiliary equations of motion can not be 
simultaneously satisfied
on the moduli space. 
While this requirement may seem rather innocuous, 
in the past it was believed that building theories with stable
dynamical supersymmetry breaking was rather difficult. 
This difficulty followed from a number of conditions 
which were thought to be requirements.
The classic ``requirements'' for supersymmetry breaking are 
\begin{itemize}
\item Chiral matter
\item A non-perturbative superpotential generated over
      the classical moduli space
\item A tree level superpotential which completely lifts
      the classical moduli space
\item $U(1)_R$ symmetry
\end{itemize}
The reasons for these ``requirements'' are reviewed below.

The Witten index, ${\rm Tr}(-1)^F$, counts the number of supersymmetric
ground states, and necessarily vanishes if supersymmetry is 
broken.  
Non-zero energy states do not contribute to the index.\cite{index}
If supersymmetry were broken in a theory which 
contained vector matter, it would then appear that completely
integrating out this sector of the theory can not change 
the index, or the conclusion
that supersymmetry is broken. 
Supersymmetry breaking by gauge dynamics therefore seems to 
require chiral representations.  
The loop hole in this argument and behavior of the  
Witten index in non-chiral theories 
is discussed in \S \ref{vector}.

Supersymmetry breaking requires a non-zero potential.
Non-perturbative gauge dynamics (specifically gaugino condensation
or a single instanton) are known to generate a non-zero 
potential over classical moduli spaces.  
The existence of such non-perturbative
dynamics therefore seems a reasonable requirement
in order to break supersymmetry. 
This puts a strict constraint on the matter content however. 
To see this, it is useful to write the full non-perturbative
superpotential as 
$W_{NP} = f( \Lambda, X_i)$ where $f$ is a holomorphic function
and $X_i$ are the (classical) gauge invariant chiral operators.  
It is convenient to assign vanishing $R$-charge to all 
chiral superfields, $R[X_i]=0$.
In this case the $R$-charge of the gauginos and matter fermions
are $R[\lambda]=1$ and $R[\psi]=-1$. 
This $R$-charge assignment is in general anomalous, 
$\partial_{\mu} j_R^{\mu} = (C / 16 \pi^2) F^a_{\mu \nu} 
\tilde{F}^{a \mu \nu}$, where 
$C \equiv C_{\lambda} - \sum_{\psi}C_{\psi}$
with ${\rm Tr}(T^a T^b) = C \delta^{ab}$. 
Now the dynamical scale is defined as the pole of the holomorphic
(one-loop) gauge $\beta$-function
\beq
\left( \Lambda \over \mu \right)^b = e^{-8 \pi^2 / g^2(\mu) + 
    i \theta}
\label{Lambdadef}
\eq
where $b$ is the $\beta$-function coefficient 
$b=3 C_{\lambda} - 2 \sum C_{\psi}$. 
Because of the anomaly, the $\theta$ term transforms under 
a $U(1)_R$ transformation as 
$\theta \rightarrow \theta + 2 C \alpha$.
The dynamical scale (\ref{Lambdadef}) then inherits an 
$R$-charge from the anomaly,
$R[\Lambda^b]=2C$. 
In the limit $\Lambda \rightarrow 0$, the non-perturbative
superpotential should vanish or reproduce the classical
moduli space.  
For a dynamical superpotential generated by gaugino
condensation or an instanton this implies $\Lambda$
must appear raised to a positive power in $W_{NP}$. 
Since $R[W]=2$ and $R[\Lambda^b]=2C$, this implies
lifting of the moduli space requires 
$\sum_{\psi} C_{\psi} < C_{\lambda}$ for an asymptotically
free theory. 
Large matter representations therefore do not generate a dynamical
superpotential, and 
were believed not to break supersymmetry.
However, as discussed in \S 3,
such theories can in fact break supersymmetry
by mechanisms other than a dynamically generated superpotential.

The magnitude of a dynamical superpotential in the ground
state depends on the expectation values on the moduli space.
Subgroups of the gauge group which are Higgsed on the moduli 
space become weaker as the expectation values increase. 
If these subgroups generate a dynamical superpotential, 
then the dynamical scale and potential 
approach zero as the expectation values go to infinity. 
If such directions on the classical moduli space 
are unlifted by a tree level superpotential
the theory exhibits run away behavior to infinite expectation
values with zero coupling and unbroken supersymmetry. 
Lifting all classical flat directions with a tree level 
superpotential therefore seems to be a reasonable requirement
for supersymmetry breaking. 
However, it is possible that along certain directions in moduli
space some fields gain mass from a tree level superpotential. 
The subgroups under which these fields transform become more
confining and more strongly coupled along these directions if 
$b_L - b_H >0$, where $b_L$ and $b_H$ are the $\beta$-function
coefficients in the high energy theory and effective 
low energy theory with the heavy matter integrated out. 
In this case the dynamical potential can grow as the
expectation values increase along these directions.
The quantum removal of directions which are classically
unlifted in theories which break supersymmetry is discussed
in \S 4.


Supersymmetry breaking requires non-vanishing auxiliary 
expectation values on the moduli space, as discussed in the previous
section. 
For $n$ fields, the vanishing of the auxiliary equations of 
motion, 
$F_{\phi}= \partial_{\phi} W=0$,
amount to $n$ equations in $n$ unknowns. 
Generically this system has a solution, and supersymmetry is
unbroken. 
However, if the theory possesses a $U(1)_R$ symmetry {\it and}
a field $\phi_i$ carrying non-zero $R$-charge has an 
expectation value, it is possible to redefine the superpotential
by\cite{Rbreaking} 
\beq
W = \phi^{2/R_i}_i W(\phi_j / \phi_i^{R_j / R_i})
\eq
The auxiliary equations of motion then become 
$n$ equations in only $n-1$ unknowns. 
Generically this system does not have a solution, and 
supersymmetry is broken. 
Note that a non-$R$-symmetry would reduce both 
the number of equations
and unknowns. 

While a spontaneously broken $U(1)_R$ symmetry generically 
implies broken
supersymmetry, this is not a necessary condition.\cite{Rbreaking}
However, almost every known example of dynamical supersymmetry
breaking has a $U(1)_R$ symmetry in some limit. 



It is now apparent that the classic ``requirements'' which are
sufficient, are certainly not necessary for supersymmetry
breaking. 
The realization that supersymmetry may be broken without
chiral matter, a dynamically generated superpotential, 
or a classical potential which completely lifts the moduli 
space has followed the recent improved understanding of strongly
coupled supersymmetric theories. 
The mechanisms by which supersymmetry may be broken without these
are detailed in \S 3 and \S 4.

\subsection{The $SU(3) \times SU(2)$ Model}

The simplest, and best studied, model which illustrates supersymmetry
breaking by the classic mechanism of a dynamically generated
superpotential is the $SU(3) \times SU(2)$ model
of Affleck, Dine, and Seiberg.\cite{ads}  
This theory satisfies all the classic requirements for supersymmetry
breaking outlined in the previous section. 

The matter content of the model is 
\beq
\bea{ccc}
        & SU(3) \times SU(2) & \\
        &             &  \\
P       & (\fund,\fund)     &   \\
L       & (1,\fund)       & \\
\overline{Q}_i & (\overline{\fund},1) & i=1,2 \\
\ea
\eq
This is just the one generation supersymmetric standard model
without hypercharge, the positron, or Higgs bosons.  
Classically, there is a moduli space parameterized by three invariants:
$Z=P^2
\overline Q_1 \overline Q_2$, 
$X_i = P L \overline Q_i$. 
The gauge group is completely broken at generic points on the moduli 
space.
At tree level there is a single
renormalizable coupling which can be added to the superpotential,
$W_{tree} = \lambda X_1$. 
This superpotential leaves invariant
non-anomalous accidental 
$U(1)_R$ and $U(1)$ flavor symmetries, and 
completely lifts the classical moduli space.  
Classically,
there is a supersymmetric ground state at the origin,
with the gauge symmetries unbroken. 

In the quantum theory 
the exact non-perturbative superpotential 
over the classical moduli space is fixed by 
holomorphy and symmetries to be 
\beq
W = { \Lambda_3^7 \over Z} + \lambda X_1
\label{ttexact}
\eq
where $\Lambda_3$ is the $SU(3)$ dynamical scale.
The dynamical term gives a potential which grows at 
small expectation values, while the tree level
terms gives a potential which grows at 
large expectation values. 
Since the model has a $U(1)_R$ symmetry, the ground state
has non-zero vacuum energy and supersymmetry is broken. 
Parametrically, for
$\lambda \ll 1$, the field expectation values and vacuum energy scale
as $\phi \sim \Lambda_3 / \lambda^{1/7}$ and $V \sim |\lambda^2 (
\Lambda_3 / \lambda^{1/7})^4| = 
|\lambda^{10/7} \Lambda_3^4|$.\cite{ads}

It is important to note that $U(1)_R$ symmetry 
plays a very important role in this model. 
It is very easy to construct models with both 
tree level and non-perturbative terms which stabilize the ground
state. 
However, without a $U(1)_R$ symmetry there is generally
an interference between various contributions to $F$-terms,
leading to $F=0$ at a finite expectation value with 
supersymmetry unbroken.

Notice that the exact superpotential (\ref{ttexact})
is independent of $\Lambda_2$. 
While it follows from holomorphy and symmetries
that the classical moduli space is unlifted by $SU(2)$
dynamics, it seems
rather odd that in the limit $\Lambda_2 \gg \Lambda_3$, the 
$SU(2)$ dynamics would have no effect on the non-supersymmetric
ground state. 
This was the original puzzle that led to the observation of 
supersymmetry breaking by the quantum deformation of the moduli 
space, as discussed in \S \ref{Qdeform}.


\section{Supersymmetry Breaking Without a Dynamical Superpotential}
\label{Wzero}

Perhaps the most important of the classic requirements for 
supersymmetry breaking is a dynamical superpotential generated 
by gaugino condensation or an instanton over the classical 
moduli space. 
In the absence of a tree level superpotential these are the only
non-perturbative effects which can lift the moduli space
and generate a potential.  
However, many other types of non-perturbative dynamics
are now understood.   
In the presence of a tree level superpotential these
can in principle give rise to additional mechanisms for supersymmetry
breaking. 
In this section the mechanisms of supersymmetry breaking 
by confinement and by the quantum deformation of the moduli space
are reviewed.  


\subsection{Confinement}

Asymptotically free gauge 
theories are necessarily strongly coupled
at the origin of moduli space, at least in 
terms of the microscopic degrees of freedom.  
However, as discussed in \S 2.1, non-perturbative
degrees of freedom often become massless at the origin of
moduli space in order to saturate global anomalies,
and represent the infrared degrees of freedom of the the
theory.
If these non-perturbative degrees of freedom do not
transform under any (magnetic) gauge symmetry, the
infrared theory is a free 
Wess-Zumino model with confined chiral
multiplets, and perhaps with superpotential couplings.
A tree level superpotential in the microscopic theory
in general leads to a modification of the effective
superpotential in the confined theory.  
Even if the microscopic tree level superpotential
has no linear terms, it is possible that
the effective superpotential is linear in a confined
field, thereby inducing a non-zero auxiliary component. 
Supersymmetry may then be broken by the O'Raifeartaigh 
mechanism in the confined theory. 
In the classical theory there is a supersymmetric ground
state at the origin, while in the confined quantum theory
the superpotential is not stationary at the origin and
supersymmetry is broken.
The mechanism for supersymmetry breaking is therefore confinement. 

The simplest example of supersymmetry breaking by confinement
is for the theory\cite{iss}
\beq
\bea{cc}
        & SU(2) \\
        &        \\
Q       & \symthree  \\
\ea
\eq
This theory has a single flat direction parameterized by 
$X=Q^4$. 
At the origin of moduli space, $X=0$, the 
gauge invariant composite $X$ has the correct global quantum 
numbers to saturate the global anomalies.  
It is therefore consistent to postulate that at the origin
$X$ is in fact a canonically normalized confined degree of 
freedom so that $K(X=0)=X^{\dagger} X$.\cite{iss} 
The lowest order term that can be added to the tree level
superpotential is $X$ itself, 
$W_{tree} = \gamma X$. 
In the classical theory this completely lifts the moduli space,
leaving a supersymmetric vacuum at the origin.  
However, in the quantum theory, with $X$ confined, 
the potential does not vanish at the origin, 
$V = \gamma^2 \Lambda_2^6$, and supersymmetry is broken. 

Almost any theory which breaks supersymmetry by a dynamically
generated superpotential can be deformed to one in which
the relevant description is tree level 
O'Raifeartaigh breaking in terms of confined degrees of freedom.
This is accomplished by 
integrating in enough light vector matter so that all the gauge 
groups confine.\cite{susymq}

\subsection{Quantum Deformation of Moduli Space}
\label{Qdeform}

The patterns of spontaneous breaking for global 
or gauge symmetries on a quantum moduli space may 
differ from those on the classical moduli space.
In this case the theory is said to have a quantum deformed
moduli space.  
For example, at the origin of a classical moduli space,
all fields have zero expectation value, and the global
symmetries are unbroken. 
However, in the quantum theory some of the global symmetries
can remain broken everywhere on the moduli space.\cite{nati} 
Points which are part of the classical moduli space can
therefore be removed by the quantum deformation. 
If tree level interactions give vanishing potential and 
auxiliary components only at points on the classical moduli 
space which are removed by quantum deformation, 
supersymmetry is broken in the quantum theory.\cite{susymq}
The mechanism for supersymmetry breaking is therefore quantum 
deformation of the moduli space. 

The mechanism of supersymmetry breaking by quantum
deformation of the moduli space is actually contained
within the $SU(3) \times SU(2)$ model discussed in \S 2.3.
To see this consider the limit 
$\Lambda_2 \gg \Lambda_3$.
In this limit the $SU(3)$ is weakly gauged at the
scale $\Lambda_2$.
Treating the $SU(3)$ as a weakly gauged global symmetry, 
the $SU(2)$ then has $N_f=2$ flavors (four fields) in 
the fundamental representation,  
namely $P$ and $L$.  
This theory has a quantum deformed moduli space.\cite{nati}
In terms of $SU(2)$ singlet moduli, it is described by
$\widehat{q} = PL/
\Lambda_2$ $\in$ $\bf 3$ of $SU(3)$ and 
$\widehat{\overline q} = P^2/ \Lambda_2$ $\in$ 
$\bf \overline{3}$ of $SU(3)$, subject to the
constraint  
$\widehat{q} \widehat{\overline q} = \Lambda_2^2$. 
Since $\widehat{q}$ and $\widehat{\overline q}$ transform under
$SU(3)$, on the quantum moduli space the $SU(3)$ is generically
completely broken. 
The maximal unbroken subgroup occurs at the point 
${\overline U} = {\overline D} =0$ and $\widehat{q} =
\widehat{\overline q} = \Lambda_2$ for which there is an unbroken
$SU(2)^{\prime} \subset SU(3)$.  
As mentioned in \S 2.3, the classical
potential vanishes only at the point for which $SU(3)$ is unbroken.
This point is removed from the quantum moduli space.
In this limit supersymmetry is therefore broken by the 
quantum deformation of the $SU(2)$ moduli space.\cite{susymq,talk}
The tree level superpotential at the maximal symmetry point
is $W = \lambda \Lambda_2^2 S_d$, where $S_d$ is 
the $SU(2)^{\prime}$
singlet component of $\overline{D}$. 
With this, quantum deformation of the 
$SU(2)$ moduli space induces an auxiliary expectation value with 
vacuum energy $V \sim |\lambda^2 \Lambda_2^4|$.  

In the limit $\Lambda_2 \gg \Lambda_3$ it is the 
non-perturbative $SU(2)$ dynamics 
which breaks supersymmetry, even though these dynamics do 
not lift the classical moduli space (c.f. Eq. (\ref{ttexact})).
This is a clear example that the exact superpotential
over the classical moduli does not always contain all the
relevant information.

\section{Supersymmetry Breaking with Vector Matter}
\label{vector}

As discussed in \S 2.2, it appears that supersymmetry
breaking by gauge dynamics requires chiral representations. 
It is however actually possible to break supersymmetry 
with vector matter.
The supersymmetric vacua which should exist in such a theory
in some sense reside at the boundary of field space where some
fields have infinite value (as shown explicitly in the example 
below).   
An unregulated calculation of ${\rm Tr}(-1)^F$ is therefore not
well defined.
But a regulated calculation, 
in which only
finite field values are weighted, 
can be suitably defined, and may vanish. 
If in fact there are no supersymmetric vacua for finite
field values, then supersymmetry is broken.


The simplest example of supersymmetry breaking with vector
matter is for the theory
\beq
\bea{ccc}
        & SU(2) & \\
        &       &  \\
Q_i       & \fund   & i=1,\dots,4 \\

S^{[ij]}  & 1     & i,j=1,\dots,4 \\
\ea
\eq
Classically there is a moduli space parameterized by 
$S^{ij}$ and $M_{ij} = Q_i Q_j$ subject to ${\rm Pf}M=0$. 
At tree level the superpotential 
$W_{tree} = \lambda S^{ij} M_{ij}$ completely lifts the
$M_{ij}$ but leaves $S^{ij}$ undetermined.
In the quantum theory, for $\lambda=0$, the $M_{ij}$ moduli 
space is deformed, and the classical constraint is modified
to ${\rm Pf}M = \Lambda_2^4$.\cite{nati}
Holomorphy and symmetries may be used to 
show that the quantum constraint is not modified for  
$\lambda \neq 0$.\cite{susymq} 
The $S^{ij}$ auxiliary equations of motions, 
$\lambda M_{ij} =0$, are then incompatible with the quantum 
constraint ${\rm Pf}M = \Lambda_2^4$. 
The classical moduli space is completely lifted for 
$\lambda \neq 0$, and supersymmetry is broken. 
The tree level superpotential 
on the quantum moduli space is\cite{susymq}  
\beq
W = \lambda M_5 S_5\pm 2 \lambda \Lambda_2^2 S_0
\eq
where $S^{ij} =\{ S_5, S_0 \}$ with 
$S_5, M_5 \in {\bf 5}$ of $SP(2)_F \subset SU(4)_F$ global
flavor symmetry, and $S_0 \in {\bf 1}$ of $SP(2)_F$.   
The first term pairs the $\widehat{M}_5 = M_5 / \Lambda_2$
and $S_5$ moduli into a massive Dirac state, while the 
second term induces an auxiliary expectation value and 
vacuum energy $V \sim |\lambda^2 \Lambda_2^4|$.
This theory demonstrates both supersymmetry
breaking with vector matter,\cite{susymq,talk,yan} and by the quantum
deformation of the moduli space.\cite{susymq,talk}

This theory has a pseudo-flat direction corresponding to the
$S_0$ component of $S^{ij}$ along which 
$V \sim |\lambda^2 \Lambda_2^4|$. 
This direction would be precisely flat if the K\"{a}hler potential
for $S$ were precisely canonical, but is lifted by quantum 
corrections.  
The index ${\rm Tr}(-1)^F$ can change 
since vacua can move continuously in or out 
from infinity along this direction under small deformations
of the theory. 
To see this consider the effective theory along the 
pseudo-flat direction $S_0$ with 
$W = \pm 2 \lambda \Lambda_2^2 S_0 + \epsilon S_0^2$. 
For $\epsilon =0$, the vacuum energy along the entire $S_0$ direction
is $V \sim | \lambda^2 \Lambda_2^4|$. 
However, for $\epsilon \neq 0$ there are two supersymmetric
ground states at $S_0 = \pm \lambda \Lambda_2^2 /\epsilon$.
For $\epsilon \rightarrow 0$ these ground states are
sent to $\infty$ along the pseudo-flat direction. 
The theory with $\epsilon =0$ (enforced by discrete or continuous
symmetries) breaks supersymmetry, while that with 
$\epsilon \neq 0$ does not. 
In this way the properly defined, regulated, 
${\rm Tr}(-1)^F$ is discontinuous at $\epsilon =0$.
The existence of a pseudo-flat direction along which the 
index can change is generic to non-chiral models of 
supersymmetry breaking.

This theory also exhibits quantum removal of classical flat
directions.\cite{susymq,talk}
The $S^{ij}$ moduli are not lifted by the tree level
superpotential, and are only lifted by quantum effects.
The existence of these classically unlifted directions contradicts
one of the classic ``requirements'' for supersymmetry breaking
discussed in \S 2.2.
In contrast to flat directions along which a
gauge group is Higgsed and becomes weaker, here the matter fields
become more massive and the theory becomes more strongly coupled,
leading to a vacuum energy which 
does not vanish even infinitely far
along the pseudo-flat direction.  
It is also possible to construct theories with classically
unlifted directions along which one subgroup
is Higgsed and becomes weaker, while another subgroup becomes more
confining and stronger, with the result that the total potential
actually grows at large expectation values.\cite{susymq}

\section{Dual Descriptions of Supersymmetry Breaking}

Identifying the relevant low energy degrees of freedom in 
models of supersymmetry breaking is important in giving
a proper description of the ground state. 
It is now understood that certain asymptotically free
gauge theories flow in the infrared to magnetic gauge
theories with different gauge symmetry and matter 
representations.\cite{sem}
In a theory with multiple scales,
if the magnetic scale is well above the supersymmetry
breaking scale, the relevant degrees in the non-supersymmetric
ground state are the infrared magnetic ones, rather than
the ultraviolet electric ones. 
By interchanging the magnetic and supersymmetry breaking
scales, the magnetic description of supersymmetry
breaking can often be continuously connected
to an electric description. 
In this way seemingly disparate models of supersymmetry
breaking can be related by duality.

Examples of dual descriptions of supersymmetry breaking 
are for the theories\cite{susydual} 
\beq
\bea{ccc}
\quad     & SU(N) \times {SP}({1 \over 2}(N-5)) & \quad \\
          &               &               \\
{A}            & (\asym,1)       & \quad         \\
{\overline{P}} & (\overline{\fund},\fund)       & \quad  \\
{ L}   & (1,\fund)   & \quad \\
\overline{U}   & (\overline{\fund},1)   & \quad \\ 
\ea
\eq
with $N \geq 11$ and odd.
These theories are just the Affleck, Dine, Seiberg, $SU(N)$ 
theories with an $\asym$ and $N-4$ 
$\overline{\fund}$,\cite{ads} with
the maximal $SP({1 \over 2}(N-5))$ flavor symmetry acting
on the $\overline{\fund}$ promoted to a gauge symmetry, and 
additional matter to cancel anomalies. 
The classical moduli space is parameterized by 
$V^k$ and  $Q V^{k-1} L$, 
$k=1,\dots,{1 \over 2}(N-5)$ where 
$V_{\alpha \beta} = A \overline{P}_{\alpha} \overline{P}_{\beta}$,
and $Q_{\alpha} = A \overline{P}_{\alpha} \overline{U}$,
and $\alpha, \beta$ are $SP(M)$ indices.
On the moduli space the gauge group is generically broken to 
$SU(5) \subset SU(N)$,
with $\asym$ and $\overline{\fund}$ of $SU(5)$ remaining. 
At tree level there is a single renormalizable coupling
which can be added to the superpotential
$W_{{tree}} = \lambda V$.
This superpotential leaves invariant a non-anomalous $U(1)_R$
symmetry, and completely lifts the classical moduli space. 
Classically there is a supersymmetric ground state at the origin.

Quantum mechanically, the non-perturbative $SU(N)$ dynamics lift the
classical supersymmetric ground state at the origin and supersymmetry
is broken.  The low energy description of supersymmetry breaking in
the ground state depends on the relative importance of the $SU(N)$ and
$SP({1 \over 2}(N-5))$ non-perturbative dynamics.  If the
$SP(M)$ is weakly coupled in the ground state, it may be
treated classically.
In this case the unbroken $SU(5)$ with 
$\asym$ and $\overline{\fund}$ generates a potential and 
breaks supersymmetry.\cite{ads}
The position of the ground state is then determined by a balance
between this dynamically generated 
potential and the tree level potential.  
This is the electric description of the theory in terms of the
underlying ultraviolet degrees of freedom.

If the $SP({1 \over 2}(N-5))$ is strongly coupled in the ground state,
its non-perturbative dynamics can not be ignored.  
For $\Lambda_{SP} \gg \Lambda_{SU}$, 
$SU(N)$ is weakly gauged at the scale
$\Lambda_{SP}$, and may be treated as a weakly gauged flavor symmetry.
The $SP({1 \over 2}(N-5))$ therefore has ${1 \over 2}(N+1)$ flavors
($N+1$ $\fund$) and for $N \geq 11$ flows in the infrared towards a
weakly coupled theory in a free magnetic phase.  
The weakly coupled
magnetic description has gauge group $SU(N) \times \widetilde{SU}(2)$ with
``mesons'' 
$\widehat{\overline{A}}=\overline{A} / \Lambda_{SP} = 
  \overline{P}^2 / \Lambda_{SP}$ $\in$
$\overline{\asym}$ of $SU(N)$ and 
$\widehat{\overline{D}} = \overline{D} / \Lambda_{SP} = 
\overline{P} L/ \Lambda_{SP}$ $\in$ $\overline{\fund}$ of
$SU(N)$, and dual ``magnetic'' quarks $\widetilde{P}$ 
$\in$ $(\fund,\fund)$ of 
$SU(N) \times \widetilde{SU}(2)$ 
and $\widetilde{L}$ $\in$ $\fund$ of $\widetilde{SU}(2)$.  
For expectation values much less than $\Lambda_{SP}$ these fields, along
with the electric fields $A$ and $\overline{U}$, make up the
canonically normalized degrees of freedom.  
The matter content of this
free magnetic phase is just 
an $SU(N) \times SU(2)$ generalization of the $SU(3) \times SU(2)$ 
model discussed in \S 2.3, 
with an
additional flavor of $\asym$ and $\overline{\asym}$ of $SU(N)$.  
This is the magnetic description of the theory.


In the absence of the electric tree level superpotential,
the moduli space of the free magnetic theory is 
parameterized by 
$Z = \widetilde{P}^2 \overline{U} {\overline{D}}$,
$X_1 = \widetilde{P} \widetilde{L} {\overline{D}}$,
$X_2 = \widetilde{P} \widetilde{L} \overline{U}$,
$\overline{V} = {\overline{A}} \widetilde{P}\widetilde{P}$,
and 
$V = {\overline{A}} A$,
subject to the dual tree level superpotential
\beq
 W_{\widetilde{tree}} = 
    { 1 \over \Lambda_{SP}} \left( \overline{V} + X_1 \right)  
\eq
This superpotential, along with the non-perturbative 
$\widetilde{SU}(2)$ dynamics ensures that the moduli space of 
the free magnetic theory coincides with the classical moduli space
of the electric theory. 
With the electric tree level superpotential,
the full tree level superpotential in the magnetic theory is 
$W_{tree} = W_{\widetilde{tree}} + \lambda V$.
It follows from symmetries, holomorphy, and limits that there
are no additional contributions to the magnetic 
tree level superpotential.   
The final term is a Dirac mass 
$m = \lambda \Lambda_{SP}$
for the pair $A$ and $\widehat{\overline{A}}$.
 For $\lambda \Lambda_{SP} \gg \Lambda_{SU}$ the Dirac
pair is much heavier than the dynamical scale in the 
free magnetic theory and may be integrated out.
Below
the scale $\lambda \Lambda_{SU}$, the effective magnetic theory is
then just the $SU(N)\times SU(2)$ theory. 
In this limit the 
$SU(N) \times SP({1 \over 2}(N-5))$ theory with an antisymmetric
representation is dual to the $SU(N) \times SU(2)$ theory with
only fundamental representations.

It is important to note that in this example
as a function of the parameters
of the microscopic theory the two descriptions
of the supersymmetry breaking ground state do not have overlapping
regions of applicability. 
In many cases duality in supersymmetry breaking models with 
product gauge groups can be used as a generator to give other
models of supersymmetry breaking.\cite{susydual}




\section*{Acknowledgments}

Most of the work reviewed in this paper was 
done in collaboration with 
K. Intriligator.

\end{document}